\documentclass[aip,reprint]{revtex4-1}
\usepackage{graphicx}
\usepackage{dcolumn}
\usepackage{bm}
\usepackage{amsmath}
\usepackage{float}
\draft 

\begin{document}

\title{Self-deformation in a dc driven helium jet micro discharge} 



\author{S. F. Xu}
\affiliation{State Key Laboratory of Advanced Optical Communication Systems and Networks, Key Laboratory for Laser Plasmas (Ministry of Education) and Department of Physics and Astronomy, Shanghai Jiao Tong University, Shanghai 200240, China}
\author{X. X. Zhong}
\email{xxzhong@sjtu.edu.cn}
\affiliation{State Key Laboratory of Advanced Optical Communication Systems and Networks, Key Laboratory for Laser Plasmas (Ministry of Education) and Department of Physics and Astronomy, Shanghai Jiao Tong University, Shanghai 200240, China}


\date{\today}

\begin{abstract}
We report on the experimental observation of three dimensional self-deformation in an atmospheric micro discharge of the helium microjet through a tube into the ambient air upon a water electrode. The geometry of the discharge system is axial symmetric. While decreasing the discharge current, three dimensional collective motion of plasma filaments are directly observed. The three dimensional configuration of the discharge self changed from an axial symmetrical horn to a rectangular horn when the water acts as a cathode.
\end{abstract}


\maketitle 

Nonlinear self formation process is one of the great interest subjects in modern natural sciences. Self-focusing, self-similar structures, and self-organized patterns have been observed and studied in nonlinear optical systems\cite{Kelley1965}, strained systems\cite{Audoly2003}, nonlinear dissipative systems\cite{Werner2015}. Nonlinear spatiotemporal self-organized patterns often can be observed in gas discharge system such as dielectric barrier discharge or dc-driven planar gas discharge system\cite{Ammelt1998,Gurlui2006}. Among them are spirals\cite{Dong2012}, concentric-ring\cite{Dong2010}, hexagonal superlattice\cite{Dong20123} and filaments\cite{Stollenwerk2006}. However, all kinds of them are two-dimensional patterns. That is to say, the nonlinear process is located on two dimensional plane.

The plasma filaments are important to the overall dynamics in a variety of plasmas both in space and in fusion laboratory experiments and are recognized as the elements to construct above mentioned complex patterns. The interaction and collective motion of filaments on two dimensional plane have been reported in many papers\cite{Muller1999,Katz2008,Dong2013}. In fact, the collective motion of filaments is the three dimensional effect of the discharge system.

\begin{figure}[H]
\center
\includegraphics{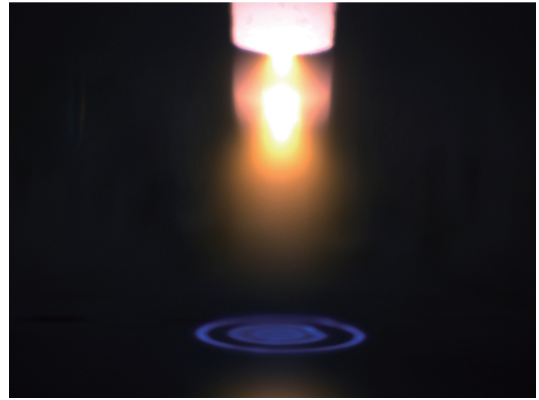}
\caption{Sample of the concentric-ring pattern observed in the helium micro discharge when water acts as an anode. The gap distance is about 3 $mm$, the helium gas flux is about 10 $sccm$ and the discharge current is about 20 $mA$.}
\label{ring}
\center
\end{figure}

Here a direct current driven atmospheric gas micro discharge system is studied in the present work, which is a three dimensional reaction-diffusion convective axial symmetric plasma system. This type of micro discharges has been used in engineering of nanoparticles\cite{Davide2010,huang2013,Ostrikov2013}. A concentric-ring pattern can be easily observed in this strongly nonlinear system when the water acts as an anode shown by Fig.~\ref{ring}. This pattern is still confined on the two dimensional discharge-water interface. However, when the water acts as cathode, instabilities of the micro discharge is charged by the plasmas filaments under weaker discharge current.

\begin{figure}[H]
\center
\includegraphics{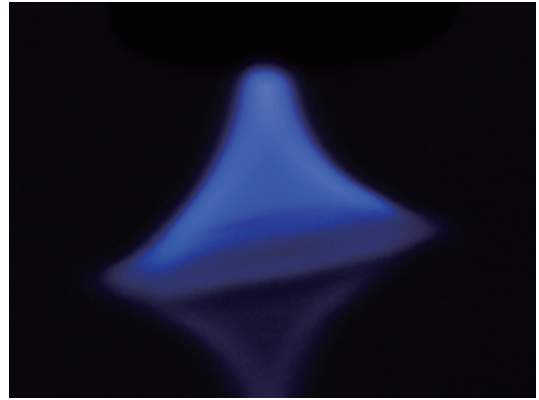}
\caption{Sample of rectangular horn observed in the micro discharge when water acts as a cathode. The gap distance is about 1 $mm$, the helium gas flux is about 10 $sccm$ and the discharge current is about 6.5 $mA$.}
\label{deform}
\center
\end{figure}

In this letter we report on the experimental observation of a three dimensional self-deformation in a helium micro discharge system shown by Fig.~\ref{deform}. The three dimensional configuration of the discharge self changed from an axial symmetrical horn to a rectangular horn.

\begin{figure}[H]
\center
\includegraphics{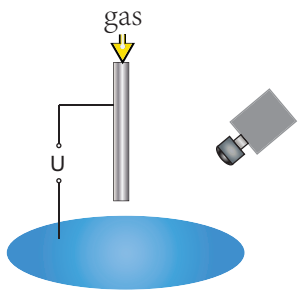}
\caption{Schematic representation of the discharge. The inner diameter of capillary is 175 $\mu m$. The helium gas flux is 10 $sccm$. A CCD camera is set in slanting view.}
\label{setup}
\center
\end{figure}

\begin{figure}[H]
\center
\includegraphics{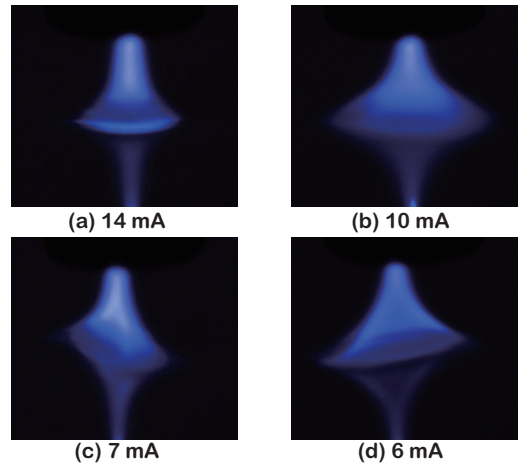}
\caption{Self-deformation process observed in the micro discharge when the water acts as a cathode. The gap distance between the exit and water surface is about 1 $mm$. The helium gas flux is about 10 $sccm$. (a) the discharge current is 14 $mA$. (b) the discharge current is 10 $mA$. (c) the discharge current is 7 $mA$. (d) the discharge current is 6 $mA$.}
\label{serie}
\center
\end{figure}

\begin{figure}[H]
\center
\includegraphics{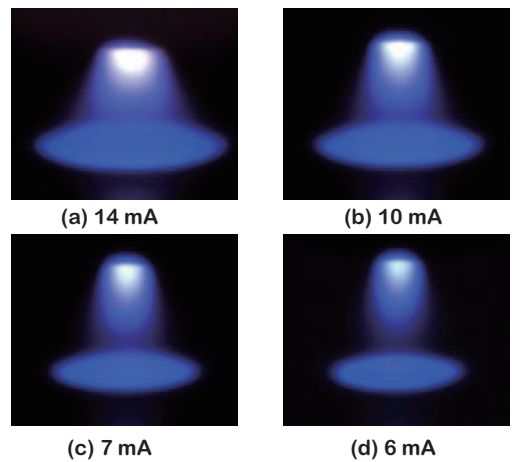}
\caption{Images of the micro discharge when the water acts as an anode. The gap distance between the exit and water surface is about 1 $mm$. The helium gas flux is about 10 $sccm$. (a) the discharge current is 14 $mA$. (b) the discharge current is 10 $mA$. (c) the discharge current is 7 $mA$. (d) the discharge current is 6 $mA$.}
\label{anode}
\center
\end{figure}

The experimental system under consideration is schematically shown in Fig.~\ref{setup}. The helium gas through a stainless-steel capillary jets into the ambient air at atmospheric pressure. The capillary acts as an electrode and its internal diameter is $d=175$ $\mu m$. The other electrode is water in a quartz dish. The water is connected to the power supply by a platinum electrode. The diameter of quartz dish is about 60 $mm$ and volume is about 25 $mL$. The resistance of water is about 1430 $\Omega$. The helium gas flux controlled by a mass flux controller (MFC) is fixed to $f=10$ $sccm$. The helium gas velocity $v=\frac{4f}{\pi d^2}$ at the exit is about 6.93 m/s and the Reynolds number $Re=\frac{v\rho d}{\mu}$ is about 10.62, where $\rho$ is the density of the helium gas and $\mu$ is the viscosity coefficient of the helium gas\cite{xu2013}. The initial gap distance between the exit of the capillary and the water surface is about 1 $mm$ adjusted by a 3D mobile platform. A dc constant-current power supply drives the micro discharge, which actually is helium and air mixture discharge due to convective diffusion. A slanting view CCD (Manta G201C) is installed to acquire micro discharge image.
A self-deformation process in the micro discharge is clearly shown in Fig.~\ref{serie}. At a larger discharge current, a stable micro glow discharge was observed when the water acts as cathode shown by Fig~\ref{serie} (a). The three dimensional structure of this micro discharge luminance is similar to that of a typical dc glow discharge. The discharge looks like a three dimensional axial symmetrical horn and is in the homogeneous mode. While decreasing the discharge current, for example to 10 $mA$, plasma filaments emerged around the previous axial symmetrical horn shown in Fig.~\ref{serie} (b). Micro plasma filaments were not static but rotated around the inner glow discharge, which showed the prominent instability of the micro plasma. Even though the discharge transited from the homogeneous mode to the mixture mode of homogeneous and filamentary modes, the micro discharge luminance was still axial symmetrical just expanded in the radial direction. But decreasing the discharge current below 7 $mA$ shown in Fig~\ref{serie} (c) and (d), the axial symmetrical structure is destroyed, the three dimensional luminance configuration of the micro discharge changed from an axial symmetrical horn to a rectangular horn. The rectangular horn has two symmetry planes of symmetry, which are much less than infinite symmetry planes of the axial symmetrical horn. A suitable comparison with the system when the water acts as an anode was shown by Fig.~\ref{anode}, where the micro discharges are in the homogeneous mode.

The mechanisms of self-deformation require further studies. As known, magnetic field has an effect on the motion of plasma filaments\cite{Schwabe2011}. Besides instabilities of helium gas flow and waves in the discharge\cite{Lu2014}, the discharge-liquid interactions also should be considered, which causes differences between the discharges when the water acts as a cathode and an anode\cite{xu2015}.

In summary, we have described experiments on the self-deformation of a dc driven micro gas jet discharge system. A three dimensional collective motion of plasmas filaments is directly observed. Decreasing symmetrical degree of freedom, the three dimensional configuration of the discharge luminance self changed from an axial symmetrical horn to a rectangular horn when the water acts as a cathode. The explanation of the self-deformation mechanism is an aim of a future research.

\begin{acknowledgments}
XXZ and SFX acknowledge support from the NSFC (Grant No. 11275127, 90923005), STCSM (Grant No. 09ZR1414600).
\end{acknowledgments}
\providecommand{\noopsort}[1]{}\providecommand{\singleletter}[1]{#1}%

\end{document}